\documentclass[aps,prb,twocolumn,showkeys,longbibliography,superscriptaddress, footinbib]{revtex4-1}

\usepackage{mathtools} 

\usepackage{xcolor}
\usepackage[normalem]{ulem} 
\usepackage{amsfonts}
\usepackage{comment}
\usepackage{amsmath}
\usepackage{physics}
\usepackage{graphicx}
\usepackage{makecell,multirow}
\usepackage{babel}
\usepackage{amstext}
\usepackage{esint}
\usepackage[unicode=true,pdfusetitle,
bookmarks=true,bookmarksnumbered=false,bookmarksopen=false, breaklinks=false,pdfborder={0 0 1},backref=false,colorlinks=false]
 {hyperref}

\makeatletter
\AtBeginDocument{\let\LS@rot\@undefined} 
\makeatother							 

\renewcommand{\vec}[1]{{\boldsymbol #1}}

\newcommand{\minus}{\scalebox{0.75}[1.0]{$-$}}

\begin{document}

\title{
Drag conductance induced by neutral-mode localization in fractional quantum Hall junctions }

\author{Jinhong Park}
\affiliation{\mbox{Institute for Quantum Materials and Technologies, Karlsruhe Institute of Technology, 76131 Karlsruhe, Germany}}
\affiliation{\mbox{Institut f{\"u}r Theorie der Kondensierten Materie, Karlsruhe Institute of Technology, 76131 Karlsruhe, Germany}}
\author{Moshe Goldstein}
\affiliation{Raymond and Beverly Sackler School of Physics and Astronomy, Tel Aviv University, Tel Aviv 6997801, Israel}
\author{Yuval Gefen}
\affiliation{Department of Condensed Matter Physics, Weizmann Institute of Science, Rehovot 76100, Israel}
\author{Alexander D. Mirlin}
\affiliation{\mbox{Institute for Quantum Materials and Technologies, Karlsruhe Institute of Technology, 76131 Karlsruhe, Germany}}
\affiliation{\mbox{Institut f{\"u}r Theorie der Kondensierten Materie, Karlsruhe Institute of Technology, 76131 Karlsruhe, Germany}}
\author{Jukka I. V\"{a}yrynen}
\affiliation{Department of Physics and Astronomy, Purdue University, West Lafayette, Indiana 47907, USA}

\date{\today}
\begin{abstract}
A junction of two 2/3 fractional quantum Hall (FQH) 
edges, with no charge tunneling between them, may exhibit Anderson localization of neutral modes.
Manifestations of such localization in transport properties of the junction are explored. There are two competing localization channels, ``neutral-mode superconductivity'' and ``neutral-mode backscattering''. Localization in any of these channels leads to an effective theory of the junction that is characteristic for FQH effect of bosons, with a minimal integer excitation charge equal to two, and with elementary quasiparticle charge equal to 2/3. These values can be measured by studying shot noise in tunneling experiments.
Under the assumption of ballistic transport in the arms connecting the junction to contacts, the two-terminal conductance of the junction is found to be 4/3 for the former localization channel and 1/3 for the latter.  
The four-terminal conductance matrix reveals in this regime a strong quantized drag between the edges induced by neutral-mode localization.  
The two localization channels lead to opposite signs of the drag conductance, equal to $\pm 1/4$,
which can also be interpreted as a special type of Andreev scattering. Coherent random tunneling in arms of the device (which are segments of 2/3 edges) leads to strong mesoscopic fluctuations of the conductance matrix. 
In the case of fully equilibrated arms, transport
via the junction is insensitive to neutral-mode localization: The two-terminal conductance is quantized to 2/3 and the drag is absent. 

\end{abstract}

\maketitle

\section{Introduction}

The fractional quantum Hall (FQH) effect~\cite{Tsui_fqh_1982, Laughlin_fqh_1983, Haldane_fqh_1983, Halperin1984statistics,Halperin_FQHE_2020} 
gives rise to a remarkable variety of topological states of matter. The topological order, as well as the associated properties of the excitations, can be detected and explored by investigations of transport properties of FQH edges in various, appropriately designed geometries~\cite{Wen_Topological_1995,Heiblum2020edge}. The primary transport observables---the electric and thermal conductances---have been intensively studied theoretically \cite{Kane_Randomness_1994,Kane_Impurity_1995,kane_quantized_1997,Rosenow_signatures_2010,Protopopov_transport_2_3_2017,Nosiglia2018,Spanslatt2021}. 
The emergent picture is particularly rich for FQH edges supporting counterpropagating modes; in that case, transport observables depend crucially on inter-mode dynamics (including presence or absence of inter-mode scattering and equilibration) determining the transport regime.
A paradigmatic example is the $\nu=2/3$ state, with the electric conductance equal to  $G = (4/3) e^2/h$ in the ballistic regime and
$G = (2/3) e^2/h$ in the inelastically equilibrated regime \cite{Protopopov_transport_2_3_2017}. 
Importantly, both these values are governed by the topology of the state, i.e., by its $K$ matrix and $\vec{t}$ vector. Specifically, $K = \text{diag} (1, -3)$ and $\vec{t}^T = (1,1)$ for the $\nu = 2/3$ state in the basis of 1 and 1/3 modes. The conductance in the equilibrated regime is $G = (\vec{t}^T K^{-1} \vec{t}) e^2/h $, while in the ballistic regime $G = (\vec{t}^T \tilde{K}^{-1} \vec{t}) e^2/h $, where $\tilde{K} = \text{diag} (1, 3)$ is obtained from the diagonal matrix $K$ by making both matrix elements positive.


On the side of experiment, there has been a major progress recently in studies of transport in FQH edges with counterpropagating modes in GaAs and graphene based devices~\cite{Banerjee_Observed_2017, Banerjee_observation_2018, Melcer_Absent_2022,Dutta_Isolated_2022,Srivastav2021May,
Srivastav2022Sep, Breton2022}. 
Moreover, platforms for engineering of ``artificial'' FQH edges at interfaces between FQH states have been developed \cite{Grivnin2014, Wang2021, Hashisaka2021, Dutta_novel_2022, Hashisaka2023,Ronen2018, Cohen2019,Wang2021}. These remarkable experimental advances motivate further theoretical work on transport in complex FQH edges and edge junctions. 

FQH edges with counterpropagating modes can be topologically unstable \cite{Haldane_Stability_1995}.
For two-mode edges, this is the case only for edges with two identical counterpropagating modes (except in the absence of charge conservation~\cite{Levin_Protected_2013}). For states with three-mode edges, there are already non-trivial examples, such as the $\nu=9/5$ state \cite{Kao_Binding_1999}. For edges with four and more modes, topological instability becomes ubiquitous. A topologically unstable edge may undergo a binding transition \cite{Kao_Binding_1999}.
In the presence of disorder-induced random tunneling between the edge modes, this takes the form of Anderson localization. As a result of localization, the number of propagating modes is reduced by two (or by an integer multiple of two). This reduces the field theory of the edge, which is characterized by a $K$ matrix and a charge vector $\vec t$ to an effective theory $(K_{\rm red}, {\vec t}_{\rm red})$ describing propagating modes. The localization is reflected in transport via the FQH edge, as was studied in Ref.~\cite{Spanslatt_binding_2023} for the $\nu=9/5$ state. Very recently, a framework for calculating the conductance of a generic FQH edge undergoing localization was developed in Ref.~\cite{yutushui2024localization}.

Here we investigate transport characteristics underlined by a specific type of Anderson localization processes at the edge. Such  processes take place at a junction, the latter referring to a section of the system where two FQH edges (that we denote A and B) may interact.
The assumption is that charge tunneling between the edges A and B is suppressed, so that the dominant localization process involves charge transfer only between the modes within each of the edges A and B,  but not between A and B. We refer to this physics as "neutral-mode localization". 

More specifically, we focus on a ``minimal'' realization of such neutral-mode localization, with both edges A and B being of $\nu=2/3$ type. Such a junction is prone to two competing localization channels (which, loosely speaking, correspond to ``neutral mode backscattering'' and ``neutral mode superconducting scattering''). We study the implications of the localization in each of these channels for charge transport via the junction in two-terminal and four-terminal setups. The four-terminal setup reveals a localization-induced drag between the edges, which can be alternatively viewed as a particular type of Andreev reflection. 

The paper is organized as follows. 
In Sec.~\ref{sec:neutral-loc}, the notion of neutral-mode localization is introduced. Applying it to a junction of two 2/3 edges, we identify two competing localization channels.
In Sec.~\ref{sec:reduced-theory}, we derive the reduced field theory describing the junction for localization in one of these channels. This yields, in particular, the elementary charges of the excitations that can be probed in tunneling experiments.  Section \ref{sec:setups} contains a discussion of transport setups analyzed in subsequent sections.
In Sec.~\ref{sec:twoterminal}, we calculate the two-terminal conductance of the junction, assuming ballistic transport in arms (2/3 edge segments) connecting the junction to the contacts. This analysis is extended in Sec.~\ref{sec:fourterminalcond}, where the four-terminal conductance matrix is determined. 
In Sec.~\ref{sec:mesoscopicfluctuation}, we include into consideration coherent random tunneling in the arms and find that it leads to strong mesoscopic fluctuations of the conductance matrix. Our results are summarized in
Sec.~\ref{sec:summary}

We set $e = \hbar = k_B = 1$ throughout the paper. The dimensional conductance $G$ is expressed in units of $e^2/h$, in accordance with the common convention.

\section{Neutral-mode localization}
\label{sec:neutral-loc}

We consider an interface between two FQH states A and B described by $(K_\text{A},\vec{t}_\text{A})$ and $(K_\text{B},\vec{t}_\text{B})$, respectively. Within Wen's $K$-matrix formalism~\cite{Wen_Topological_1995}, the resulting FQH edge junction is 
described by the effective action 
\begin{align}
\label{eq.L0}
    S_0 = \frac{1}{4\pi}\int dt dx\; \sum_{a,b=1}^d\partial_x\phi_a( K_{ab}\partial_t\phi_b-V_{ab}\partial_x\phi_b),
\end{align}
with a block-diagonal $K$-matrix and a combined $\vec t$-vector, 
\begin{align}
    K_\text{AB} = \begin{pmatrix}
        K_\text{A}&0\\0&-K_\text{B}
    \end{pmatrix},\qquad
    \vec{t}_\text{AB} = 
    \begin{pmatrix}
\vec{t}_\text{A} \\ \vec{t}_\text{B}
    \end{pmatrix} .
\end{align}
In Eq.~\eqref{eq.L0},  $V_{ab}$ is a non-universal positive-definite matrix of mode velocities and inter-mode interactions. At variance with $K$, the matrix $V$ is not block-diagonal due to inter-edge interaction. Upon canonical quantization, Eq.~\eqref{eq.L0} implies the commutation relations
\begin{align}
\label{eq.commutation}
    [\rho_a(x),\phi_{b}(x')] = i K^{-1}_{ab}\delta(x-x'),
\end{align}
where $\rho_a= \partial_x\phi_a/2\pi$ is the particle density of the mode $a$; The corresponding charge density is $t_a \rho_a$.

In the presence of disorder, processes of random tunneling between the modes are possible \cite{Moore_Classification_1997}. Such processes are described by terms in the  Hamiltonian of the form 
\begin{align} 
\label{eq:tunnelinglocalization}
    H_{\rm tun} = \int dx  \, g (x) \text{cos} (\vec{M}^T \vec{\phi} (x) + \zeta (x)) \,,
\end{align}
where $\xi (x) = g (x) e^{i \zeta (x)}$ is the disorder amplitude (with magnitude $g (x)$ and phase $\zeta (x)$). The vectors  $\vec{M}$ are integer-valued and satisfy a charge-conservation condition
\begin{equation}
Q_\vec M = 0 \,,
\label{eq:charge-conserv}
\end{equation}
where the charge $Q_\vec M$ is given by 
\begin{equation}
Q_{\vec M} = t_aK^{-1}_{ab}M_b \equiv {\vec t}^T K^{-1} {\vec M} \,,
\end{equation}
For a junction of edges A and B, we have $\vec M =(\vec M_\text{A}, \vec M_\text{B})$, and
the charge conservation condition \eqref{eq:charge-conserv}
can be rewritten as $Q_{\vec M_\text{A}} = - Q_{\vec M_\text{B}}$.
If the two edges, A and B, belong to different FQH systems and thus are separated by a vacuum, only electrons can tunnel between them, which implies an additional constraint: 
$\vec M_A = K_A \vec L_A$  and $\vec M_B = K_B \vec L_B$, where $L_A$ and $L_B$ are  
integer-valued vectors.

We note that Ref.~\cite{Levin_Protected_2013} relaxed the charge-conservation condition 
\eqref{eq:charge-conserv} because it
 considered a FQH edge coupled to a superconductor. Here, we assume that charge in the edge junction is conserved, so that Eq.~\eqref{eq:charge-conserv} must hold.

The tunneling process \eqref{eq:tunnelinglocalization} characterized by a vector $\vec M$ may induce localization if $\vec{M}$ satisfies, in addition to the charge-conservation \eqref{eq:charge-conserv}, Haldane's null-vector condition \cite{Haldane_Stability_1995}, given by \footnote{In Ref.~\cite{Haldane_Stability_1995}, Haldane's null vector condition has the form $\vec{m}^T K \vec{m} = 0$, with a local (fermionic or bosonic) field $\exp(-i \vec{m}^T \vec{\varphi})$. The field 
$\vec{\varphi}$ in Ref.~\cite{Haldane_Stability_1995} is related to our field $\vec{\phi}$ via $\vec{\varphi} = K\vec{\phi}$ (as can be seen by comparing the commutation relations). Correspondingly, equating 
$\exp(-i \vec{m}^T \vec{\varphi}) = \exp(-i \vec{M}^T \vec{\phi})$, we obtain the identification of vectors $\vec{M} = K \vec{m}$.
With this identification, Haldane's condition 
$\vec{m}^T K \vec{m} = 0$ is equivalent to our condition \eqref{eq:nullvector}.}
\begin{align} 
\label{eq:nullvector} 
    \vec{M}^T K^{-1} \vec{M} = 0\,.
\end{align}
For the localization to be operative,
the tunneling \eqref{eq:tunnelinglocalization} should be relevant in the renormalization-group (RG) sense. For any null vector $\vec{M}$, this is indeed the case in a certain region of parameters of the interaction matrix $V$. A general analysis of transport in FQH edges with localization was carried out in Ref.~\cite{yutushui2024localization}.


Here, we consider the case when the vector $\vec M$ governing the localization corresponds to no charge transfer between the edges A and B, i.e.,
\begin{equation}
Q_{{\vec M}_\text{A}}=Q_{{\vec M}_\text{B}} = 0 \,.
\label{eq:neutral-loc}
\end{equation}
We term this case ``neutral-mode localization''.  
Processes involving zero charge transfer between the edges may be dominant for several reasons. First, the neutral-mode localization can be induced by the 
inter-edge interaction, given by 
\begin{align} \label{eq:interedgeinteraction}
    H = \frac{1}{2} \int d \vec{r} d\vec{r'} \rho_{\text A} (\vec{r}) V (\vec{r} - \vec{r'}) \rho_{\text B} (\vec{r'})\,.
\end{align}
Here the local particle density operator $\rho_{ \text{A}/\text{B}}$ can be written as
\begin{align}
    \rho_{i= \text{A}/\text{B}} = \sum_{a} \rho_{a, i} + \sum_{a\neq b} \psi_{a, i}^\dagger \psi_{b, i}\,, 
\end{align}
where $\psi_{a,i }$ ($\psi_{a,i}^{\dagger}$) is the electron annihilation (creation) operator of mode $a$ on edge $i = \text{A}/\text{B}$. Following the bosonization of the electron operators, the inter-edge interaction~\eqref{eq:interedgeinteraction} has the form of \eqref{eq:tunnelinglocalization} and satisfies the condition~\eqref{eq:neutral-loc} for the neutral-mode localization. Generically, $V (\vec{r} - \vec{r}')$ is a power-law function of distance, and thus the inter-edge interaction may remain sufficiently strong for realistic setups. By contrast, charge tunneling between edges drops exponentially with the distance, thus quickly becoming negligible.
Second, one can engineer edge modes at an interface that have opposite spin polarizations~\cite{Jukka2022}. This can be achieved by using a double-quantum-well structure~\cite{Ronen2018, Cohen2019}. 
In this situation, charge tunneling will necessarily involve spin flip and thus may be strongly suppressed. 

We focus on a specific realization of neutral-mode localization: an interface of 
spin-polarized $\nu = 2/3$ FQH regions separated by a ``vacuum'' strip. Thus, each of the edges A and B is a 2/3 FQH edge consisting of counterpropagating 1 and 1/3 modes, which is described by the following $K$-matrix and $\vec{t}$-vector:
\begin{align}
    K_A = K_B = K_0 = \begin{pmatrix} 1 & 0 \\ 0 & -3 \end{pmatrix}\,, \quad 
    \vec{t}_A = \vec{t}_B = \vec{t}_0 = \begin{pmatrix}
        1 \\ 1 
    \end{pmatrix}\,.
    \label{eq:K0}
\end{align}
For this edge junction, there are two vectors
$\vec M$ that satisfy the above conditions  
\eqref{eq:charge-conserv},
\eqref{eq:nullvector}, and  
\eqref{eq:neutral-loc}, namely 
\begin{align}
\vec{M}_{\text{back}} = (1, 3, 1, 3)^T
    \label{eq:M-back}
\end{align}
and 
\begin{align}
    \vec{M}_{\text{sup}} = (1, 3, -1, -3)^T \,.
        \label{eq:M-sup}
\end{align}
 The random intermode tunneling in each of these two channels can therefore induce neutral-mode localization. The tunneling process described by
$\vec{M}_{\text{back}}$ can be viewed as backscattering of neutral excitations, while the 
process $\vec{M}_{\text{sup}}$ can be understood as a ``superconducting'' neutral-mode tunneling between the edges A and B \cite{Jukka2022}. This motivates the subscripts that we use to label these two processes.

Importantly, these two null vectors are competing, which follows from the fact that $\vec{M}_{\text{sup}}^T K^{-1} \vec{M}_{\text{back}} \neq 0$. 
Thus, once one of these null vectors drives the localization, so that the corresponding bosonic field in the argument of cosine in Eq.~\eqref{eq:tunnelinglocalization} is pinned to a minimum, the bosonic field described by the other null vector is fluctuating in position and time.  Which of the null vectors wins depends on their bare strengths and RG scaling exponents, where the latter generically depend on the interaction matrix $V$. An RG analysis in a certain subspace of matrices $V$, and with an assumption of correlated disorder, has been carried out in Ref.~\cite{Jukka2022}. 

We emphasize that the localization induced by either $\vec{M}_{\text{back}}$ or $\vec{M}_{\text{sup}}$ 
is only partial, in the sense that, out of four modes, two are localized while two propagating modes remain. We will derive a reduced theory for the propagating modes in the next section.
Clearly, a junction of two 2/3 edges is in principle fully unstable topologically, i.e., all four modes can get localized if one allows all possible tunneling processes. However, as discussed above, we allow only processes without inter-edge charge transfer, Eq.~\eqref{eq:neutral-loc}, and the reduced two-mode theory is stable with respect to such processes. 

The neutral-mode localization can be also considered in the case when both 2/3 edges A and B belong to the same FQH system and thus are separated by a strip of FQH bulk. Our analysis below for the conductances of the junction equally applies to this case. 

While, in this paper, we will focus on the neutral-mode localization at the junction of two spin-polarized $\nu = 2/3$ edges, such a localization may also take place for a  junction of two {\it spin-unpolarized} $\nu = 2/3$ edges, where neutral modes carry spin. Note that if spin is conserved, only the $\vec{M}_{\text{back}}$ localization channel~\eqref{eq:M-back} is operative. 

It is also worth  noting that an artificial FQH edge at an interface between the spin-polarized and spin-unpolarized 2/3 states has been experimentally realized in Ref.~\cite{Wang2021} and theoretically studied in Refs.~\cite{ponomarenko2023unusual,yutushui2024localization}. This setup is essentially different from the one considered in the present paper, since it crucially involves a charge tunneling between the 2/3 edges. 

\section{Reduced theory}
\label{sec:reduced-theory}

We now derive a reduced theory of the junction after neutral-mode localization induced by either \eqref{eq:M-back} or \eqref{eq:M-sup}. A general procedure for determining the reduced theory  of an edge undergoing localization was presented in Ref.~\cite{yutushui2024localization}; see, in particular, Appendix D there. 
Applying it to our model, we find that
the localization in any of the two channels
\eqref{eq:M-back} or \eqref{eq:M-sup} 
results in the same reduced theory 
\begin{align}
    K_{\text{red}} = \begin{pmatrix} 0 & 3 \\ 3& 0 \end{pmatrix}\,, \quad \vec{t}_{\text{red}} =  \begin{pmatrix} 2 \\ 0  \end{pmatrix}\,.
\label{eq:K-t-reduced}
\end{align}
The basis vectors that span the two-dimensional lattice of allowed excitations $\vec m$ of the reduced theory are, in terms of the original theory,
\begin{equation}
\label{eq:e-red-sup}
 \vec{e}_1^{\text{red}} = (1, 2, 0, -1)^T \,, \qquad \vec{e}_2^{\text{red}} = (0, -1, 0, 1)^T
 \end{equation}
 for the null vector $\vec{M}_{\text{sup}}$, and 
 \begin{equation}
 \label{eq:e-red-back}
 \vec{e}_1^{\text{red}} = (0, -1, 0, -1)^T\,, \qquad \vec{e}_2^{\text{red}} = (1, 2, 0, 1)^T
 \end{equation} 
 for $\vec{M}_{\text{back}}$.The theory \eqref{eq:K-t-reduced} is written in this basis.

 Importantly, the edge theory with $(K_{\text{red}}, \vec{t}_{\text{red}})$ as given by Eq.~\eqref{eq:K-t-reduced} is bosonic; all diagonal elements of $K_{\text{red}}$ are even and the components of $\vec{t}_{\text{red}}$ are all even. 
A generic excitation described by an integer-valued vector $\vec{m}^T = (m_1, m_2)$ in the reduced theory has charge and statistics given by 
\begin{align}
    Q_{\vec{m}} &\equiv \vec{m}^T K_{\text{red}}^{-1} \vec{t}_{\text{red}}= \frac{2}{3} m_2 \,, \nonumber \\ 
    \theta_{\vec{m}} &\equiv \pi  \vec{m}^T K_{\text{red}}^{-1}   \vec{m} = \frac{2\pi}{3} m_1 m_2\,.
\end{align}
For charged excitations, the minimal charge corresponds to $m_2 = 1$ that yields $Q_\vec{m} = 2/3$. This is twice the charge of the elementary excitation of the original theory. Furthermore, the minimal integer charge of an excitation corresponds to $m_2 = 3$ and is $Q_\vec{m} = 2$. All integer-charge excitations (that correspond to $\vec m = K_{\text{red}} \vec{l}$ with an integer-valued vector $\vec l$) have an even charge $Q_\vec{m}$ and bosonic statistics ($\theta_{\vec{m}}$ is an integer-multiple of $2\pi$). The doubling of the charge as a result of localization can be observed by measuring the shot noise in experiments on tunneling to the edge junction. For tunneling through the $\nu =2/3$ FQH liquid, the elementary charge is now 2/3 (instead of 1/3 in the absence of localization), and for tunneling via the vacuum it is 2  (instead of 1 in the absence of localization). 

We turn now to the analysis of manifestations of localization in conductances of the edge junction. We begin this analysis by discussing possible experimental transport setups in Sec.~\ref{sec:setups}.

\section{Transport setups}
\label{sec:setups}

\begin{figure}[t!]
\includegraphics[width =\columnwidth]{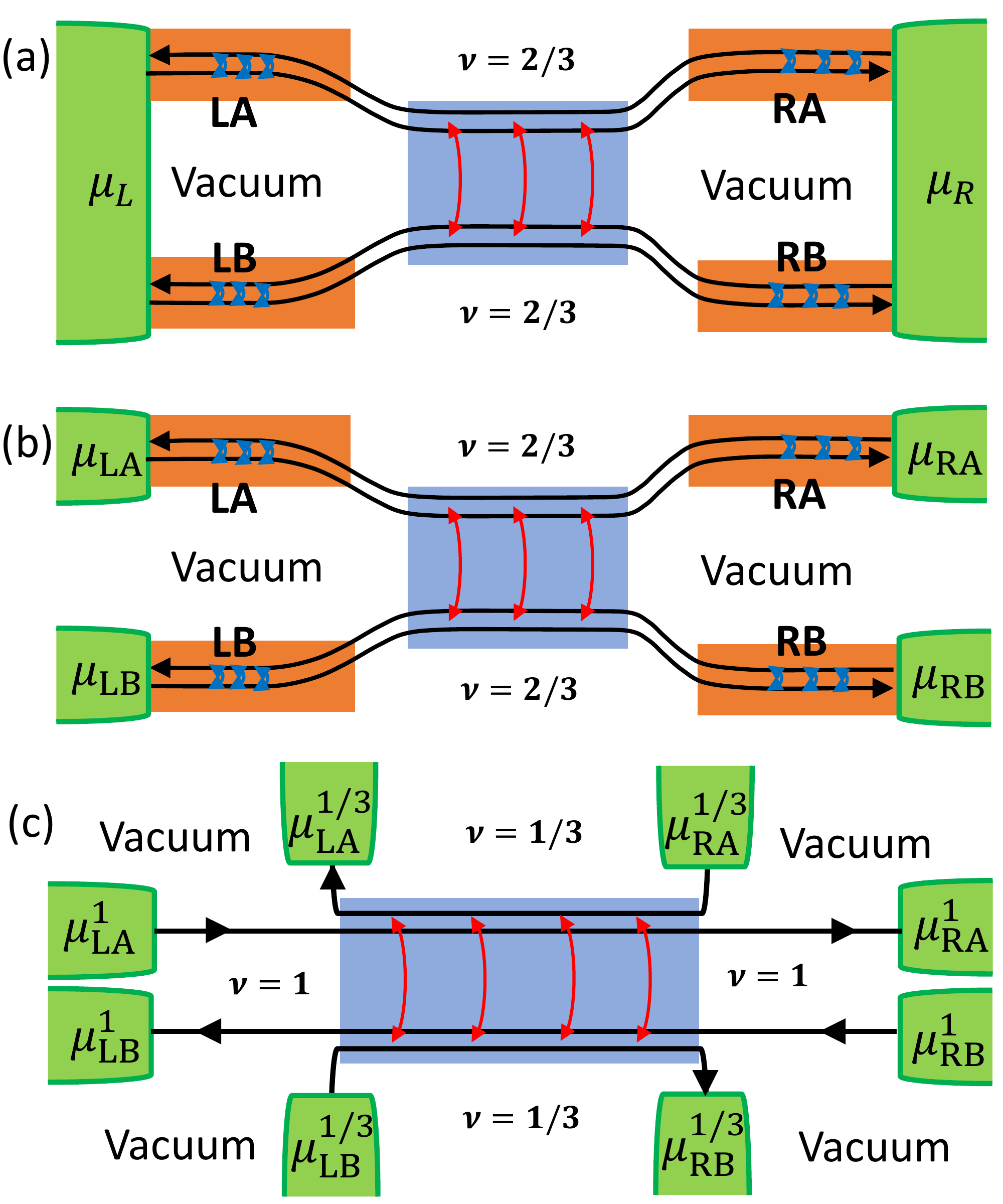}
\caption{
{\bf Schematic setups for studying transport through a junction of two 2/3 edges} (depicted as a shaded blue region). In edge A (the upper one), mode 1 propagates to the right and mode 1/3 to the left; in edge B the directions are reversed.
Localization occurs in the junction (blue central region), either in the  neutral back-scattering channel $\vec{M}_{\text{back}}$ or in the  neutral superconducting channel $\vec{M}_{\text{sup}}$. 
{\bf (a)} Two-terminal setup, with four arms LA, LB, RA, and RB that are segments  of the 2/3 edges that connect the junction to the metallic contact regions (depicted as green regions). In general, the transport through these segments may be either coherent (which in turn is subdivided into ballistic and involving random tunneling leading to mesoscopic fluctuations), or equilibrated (incoherent). {\bf (b)} Four-terminal setup with analogous arm segments LA, LB, RA, and RB.  {\bf (c)} Modification of four-terminal setup that does not contain segments of 2/3 edges in the arms. 
} 
\label{Fig:Setup}
\end{figure}

In Fig.~\ref{Fig:Setup}, we show setups that can be used in transport experiments to explore signatures of neutral-mode localization in the junction.
Figure \ref{Fig:Setup}(a) presents a two-terminal conductance setup, which involves two contacts (metallic electrodes) at electrochemical potentials $\mu_L$ and $\mu_R$. The junction is connected to the contact regions by four arms representing segments of the 2/3 edges. 
The localization (driven either  by the neutral backscattering term $\vec{M}_{\text{back}}$ or by the neutral superconducting coupling term $\vec{M}_{\text{sup}}$, as discussed above) takes place in the junction. In the arm segments, intra-edge tunneling might take place [which corresponds to $\vec{M}_{\text{intra}, \text{A}} = (1, 3, 0, 0)^T$ for the edge A and to $\vec{M}_{\text{intra}, \text{B}} = (0, 0, 1, 3)^T$ for the edge B]. These intra-edge tunneling processes do not satisfy the null vector condition \eqref{eq:nullvector} and thus cannot induce localization. At the same time, they may lead, in the coherent regime, to mesoscopic fluctuations or, at sufficiently high temperatures, to incoherent equilibration. Below, we will analyze how the conductance is influenced by different transport regimes in the arm segments.

A four-terminal version of the transport setup is depicted in
Fig.~\ref{Fig:Setup}(b). This setup allows one to independently control or measure voltages and currents in four electrodes
(with potentials $\mu_{\text{LA}}$, $\mu_{\text{RA}}$,
$\mu_{\text{LB}}$, and $\mu_{\text{RB}}$),
which provides much more information about transport in the junction. In particular, this setups makes it possible to explore how a current driven through, say, edge A induces current (or voltage) in edge B;  the effect that can, has a character of drag and can also be interpreted in terms of crossed Andreev reflection. Finally, in  Fig.~\ref{Fig:Setup}(c) we propose an alternative realization of the four terminal setup, with the 2/3 edges in the junction are artificially engineered from counterpropagating 1 and 1/3 edges. Such an artificial 2/3 edge has been already realized in Refs.~\cite{Cohen2019, Hashisaka2021, Hashisaka2023}
In this setup, $\mu_{\text{LA}}^1$, $\mu_{\text{RA}}^{1/3}$,
$\mu_{\text{LB}}^{1/3}$, and $\mu_{\text{RB}}^1$ play the same role as 
$\mu_{\text{LA}}$, $\mu_{\text{RA}}$,
$\mu_{\text{LB}}$, and $\mu_{\text{RB}}$ in
Fig.~\ref{Fig:Setup}(b). The difference is that the setup of Fig.~\ref{Fig:Setup}(c) does not involve segments of 2/3 edges in the arms that affect the conductance if tunneling there is operative.

\section{Two-terminal conductance} \label{sec:twoterminal}

We now proceed by calculating the two-terminal conductance in the setup depicted in Fig.~\ref{Fig:Setup}(a) for each of the competing localization channels [i.e., the null vectors $\vec{M}$,  see Eqs.~\eqref{eq:M-back} and  \eqref{eq:M-back}]
 governing the neutral-mode localization in the junction. In general, the value of the conductance depends not only on the localization channel in the junction, but also on the transport regimes in the arms connected to the junction. We consider here two limiting transport regimes in the segments LA, LB, RA, and RB:
(i) ballistic transport (tunneling not operative in the arms) and (ii) the fully equilibrated regime, where the length of the arms $L_{\text{arm}}$ is much larger than the inelastic equilibration length $\ell_{\text{eq}}$. In Sec.~\ref{sec:mesoscopicfluctuation} below, we supplement the present analysis by considering an additional
important regime involving coherent random tunneling in the arms.

 According to Ref.~\cite{yutushui2024localization}, the two-terminal conductance of the junction reads
\begin{align} 
\label{eq:twoterminalcondlocalization}
    G = \frac{1}{2} \big (\sum_{a, b =1}^{2} t_\text{red}^a (W^T U^T U W)^{-1}_{ab} t_{\text{red}}^b - B^2 / C \big )\,, 
\end{align}
with
\begin{align}
 B &= \vec{t}_{\text{red}} (U^T U W)^{-1} \vec{M} \,, \\ 
  C &= \vec{M}^T (U^T U)^{-1}\vec{M}\,.
\end{align}
Here $\vec M$ is the null vector governing the localization process, i.e., either
$\vec{M}_{\text{back}}$ or $\vec{M}_{\text{sup}}$, while 
$W\in \text{SL}(4,\mathbb{Z})$ is a transformation that brings $K$-matrix of the original theory to a block-diagonal form, separating the sector that undergoes localization from the remaining sector $(K_{\rm red}, {\vec t}_{\rm red})$, see Eq.~\eqref{eq:K-t-reduced}.
Further, the matrix $U$ has the form $U = O R_K$, where the matrix $R_K$~\footnote{The matrix $R_K$ was denoted by $M_K$ in Ref.~\cite{yutushui2024localization}. In the present paper, we use a different notation to avoid a possible confusion with null vector $\vec{M}$.} brings $K$ to the form $K = R_K^T \Lambda R_K$ with $\Lambda = \text{diag} (1, -1, -1, 1)$, i.e., 
\begin{align}
    R_K = \text{diag} (1, \sqrt{3}, 1, \sqrt{3}).
\end{align}
The matrix $O \in {\rm SO}(2,2)$
depends on the degree of equilibration in the edge segments LA, LB, RA, and RB.  
It is parametrized by two ``angles'' $\vartheta_{A}$ and $\vartheta_{B}$ as 
\begin{align}
    O = \begin{pmatrix} \cosh \vartheta_{A} & \sinh \vartheta_{A}  \\ \sinh \vartheta_{A} & \cosh \vartheta_{A} \end{pmatrix} \oplus
    \begin{pmatrix} \cosh \vartheta_{B} & \sinh \vartheta_{B}  \\ \sinh \vartheta_{B} & \cosh \vartheta_{B} \end{pmatrix}. 
\end{align}

Note that Eq.~\eqref{eq:twoterminalcondlocalization}
contains an additional overall factor of $1/2$
in comparison with the general formula (70) in Ref.~\cite{yutushui2024localization}.
This is because we consider here the conductance of a single junction, while 
Eq.~(70) of Ref.~\cite{yutushui2024localization}
was derived for a two-terminal conductance of a structure including two composite edges (which would mean two junctions in the present case). It should also be emphasized that, since the total filling factor of a junction considered here is $2/3 - 2/3 = 0$, there is no anomaly in the current, i.e., the current from the contact L to the contact R through the junction is 
\begin{equation}
J = (\mu_L - \mu_R) \, \frac{G}{2\pi} \,.
\end{equation}
The factor $1/2\pi$ here is due to the fact that $G$ is defined as measured in units of $e^2/h$ and we set $e = 1$ and $\hbar = 1$. 

For the case of localization in the 
$\vec{M}_{\text{sup}}$ channel, we obtain from
Eq.~\eqref{eq:twoterminalcondlocalization}:
\begin{align}
    G_{\text{sup}} &=  \Big (\frac{1 + 7 \cosh[2(\vartheta_{A} + \vartheta_{B})] - 4 \sqrt{3} \sinh [2(\vartheta_{A} + \vartheta_{B})] }{ 6\cosh(\vartheta_{A} + \vartheta_{B}) - 3 \sqrt{3}\sinh(\vartheta_{A} + \vartheta_{B}) } \Big ) \, \nonumber \\ 
    & \qquad  \times 
    \frac{1}{\cosh(\vartheta_{A} - \vartheta_{B})} 
\end{align}
In the same way, we get for the case when the localization develops in the $\vec{M}_{\text{back}}$ channel:
\begin{align}
    G_{\text{back}} =  \frac{2\cosh[2(\vartheta_{A} - \vartheta_{B})] }{ 6\cosh(\vartheta_{A} + \vartheta_{B}) - 3 \sqrt{3}\sinh(\vartheta_{A} + \vartheta_{B}) }. 
\end{align} 
We next discuss these two-terminal conductance formulas in the two limiting cases of transport through the connecting arms mentioned above. The angles $\vartheta_{A}$ and $\vartheta_{B}$ can be determined by diagonalizing the equilibration matrix $\Upsilon_{\text{AB}}$,
\begin{align} \label{eq:equalibrationmatrix}
    \Upsilon_{\text{AB}} &= \Upsilon_{\text{A}} \oplus \Upsilon_{\text{B}}\,, \nonumber \\ 
    (\Upsilon_{\text{A}/\text{B}})_{ab} &= \delta_{ab} \left(\frac{\Gamma_{a}}{t_a}+ \sum_{c} t_a^{-1} \gamma_{ac} t_c \right)  - \gamma_{ab}\,, 
\end{align}
with the matrix $U$, i.e., by finding $\vartheta_{A}$ and $\vartheta_{B}$ such that  $U \Upsilon_{\text{AB}} U^T$ is diagonal. Here $\gamma_{ab}$ are inter-mode tunneling rates and $\Gamma_a$ are tunneling rates to metallic contacts; see Appendix A in Ref.~\cite{yutushui2024localization} for more details.

When the the transport in the arms is ballistic, $\gamma_{ab}$ in Eq.~\eqref{eq:equalibrationmatrix} vanish and thus
$\vartheta_A = \vartheta_B = 0$. It yields 
\begin{align}
    G_{\text{sup}} = \frac{4}{3} \,; \qquad 
    G_{\text{back}} = \frac13 \,.
    \label{eq:G-teo-term-sup-back}
\end{align} 
We thus see that, under the assumption of ballistic transport in the arms, the conductance $G$ clearly distinguishes between the two localization channels. We also note that the value $G_{\text{sup}} = 4/3$ is the same as the ballistic conductance in the absence of any localization \cite{Protopopov_transport_2_3_2017}, i.e., the neutral-mode localization in the ``superconducting'' channel does not affect the two-terminal ballistic conductance $G=4/3$. On the other hand, the localization in the ``neutral backscattering'' channel reduces the conductance down to $G=1/3$. 

Since the localization in the ``neutral-superconducting'' channel does not manifest itself in the two-terminal conductance, one can ask whether it affects at all transport through the junction.  The answer is yes, as we will show in
in Sec.~\ref{sec:fourterminalcond} where the four-terminal terminal conductance will be studied. 

For the case of inelastic equilibration in the arms, $\gamma_{ab} \gg \Gamma_a$ in Eq.~\eqref{eq:equalibrationmatrix}. In this limit, we find $\vartheta_A = \vartheta_B =  \frac{1}{2} \log (2 + \sqrt{3})$, resulting in the conductance
\begin{align} 
\label{eq:supercondequilibration}
   G_{\text{sup}} = G_{\text{back}} = \frac{2}{3} \,,
\end{align}
The value of the conductance is thus the same, $G=2/3$ as in the equilibrated regime without any coupling of the two 2/3 edges 
\cite{Protopopov_transport_2_3_2017,Nosiglia2018}. Thus, the neutral-mode localization has no signatures in the conductance in the regime of strong inelastic equilibration in the arms. We will show in Sec.~\eqref{sec:fourterminalcond} that this applies also to the four-terminal conductance matrix. In order to eliminate (or, at least, strongly suppress) the effect of inelastic equilibartion in the arms, one can use the setup shown in
Fig.~\ref{Fig:Setup}(c). One can study the two-terminal conductance in this setup by, e.g., applying voltages $\mu_{\text{LA}}^{1} = \mu_{\text{LB}}^{1/3}$  and grounding other contacts.

\section{Four-terminal conductance} \label{sec:fourterminalcond}

We turn now to the analysis of the conductance in a four-terminal setup, see Fig.~\ref{Fig:Setup}(b), which provides more information about transport through the junction. For this purpose, we extend the formalism of Ref.~\cite{yutushui2024localization} to the calculation of four-terminal conductance.

We denote by $\vec{I}_{\ell}^{\text{jun}}$ 
with $\ell = \text{LA}, \text{LB}, \text{RA}, \text{RB}$ the two-component particle-current vectors that flow in the junction (or out of the junction) via the corresponding arms. 
 The first and second components of  $\vec{I}_{\ell}^{\text{jun}}$ represent the particle currents in the $\nu = 1$ and $\nu = 1/3$ modes, respectively. The sign choice for $\vec{I}_{\ell}^{\text{jun}}$ is such that these currents are positive when they flow in the positive direction of $x$ axis (i.e., from left to right in the figures). 
It is convenient to write these currents in terms of the respective chemical potentials $\vec\mu_{\ell}^{\text{jun}}$ defined via
\begin{align}
    \vec{I}_{\text{LA}}^{\text{jun}} &= \frac{1}{2\pi} K^{-1}_{0} \vec{\mu}_{\text{LA}}^{\text{jun}}\,, \quad 
     \vec{I}_{\text{LB}}^{\text{jun}} = -\frac{1}{2\pi} K^{-1}_{0} \vec{\mu}_{\text{LB}}^{\text{jun}}\,, \nonumber \\ 
       \vec{I}_{\text{RA}}^{\text{jun}} &= \frac{1}{2\pi} K^{-1}_{0} \vec{\mu}_{\text{RA}}^{\text{jun}}\,, \quad 
     \vec{I}_{\text{RB}}^{\text{jun}} = -\frac{1}{2\pi} K^{-1}_{0} \vec{\mu}_{\text{RB}}^{\text{jun}}\,. 
\end{align}

As shown in Ref.~\cite{yutushui2024localization}, the localization in channel $\vec M$ in the junction implies that  $\vec{\mu}_{\ell}^{\text{jun}}$ can be written in the form
\begin{align} 
\label{eq:chemicalpotentialconst}
    \begin{pmatrix}
        \vec{\mu}_{\text{LA}}^{\text{jun}} \\ \vec{\mu}_{\text{LB}}^{\text{jun}} 
    \end{pmatrix}  
    = \sum_{a=1}^{2} \mu^{\text{red}}_{a} \vec{e}_{a}^{\text{red}} + \mu_{L}^{\rm loc} \vec{M}\,, \nonumber \\
      \begin{pmatrix}
        \vec{\mu}_{\text{RA}}^{\text{jun}} \\ \vec{\mu}_{\text{RB}}^{\text{jun}} 
    \end{pmatrix}  
    = \sum_{a=1}^{2} \mu^{\text{red}}_{a} \vec{e}_{a}^{\text{red}} + \mu_{R}^{\rm loc} \vec{M}\,.
\end{align}
Here  $\vec{e}_{a=1,2}^{\text{red}}$ are two basis vectors of the reduced theory (see Eqs.
\eqref{eq:e-red-sup} and \eqref{eq:e-red-back}
for their explicit form), $\mu^{\text{red}}_a$ are the corresponding chemical potentials, while $\mu^{\rm loc}_R$ and $\mu^{\rm loc}_L$ are the chemical potentials of the localized modes.
Note that while $\{\vec{\mu}_{\ell}^{\text{jun}}\}$ contain eight (scalar) parameters, they are expressed in Eq.~\eqref{eq:chemicalpotentialconst} in terms of four chemical-potential variables. 
The reason for this is that the derivation of Eq.~\eqref{eq:chemicalpotentialconst} used four constraints following from the localization in the junction. These are two conditions 
\begin{equation}
\vec{M}^T (\vec{I}_{\rm LA}^{\text{jun}}, \vec{I}_{\rm LB}^{\text{jun}})  = 0 \,, \qquad \vec{M}^T (\vec{I}_{\rm RA}^{\text{jun}}, \vec{I}_{\rm RB}^{\text{jun}})  = 0
\label{eq:loc-condition-M}
\end{equation}
expressing the localization in channel $\vec M$ and two conditions expressing continuity of the remaining modes.

Further, the currents in the arms in the direct vicinity of metallic contacts (depicted as green regions) can be expressed as~\cite{yutushui2024localization} 
\begin{align}
   \vec{I}^{\text{con}}_{\text{LA}} &= \frac{\mu_{\text{LA}}}{2\pi} K_{0}^{-1} \vec{t}_0 +  R_{K_0}^{-1} P_{1/3} \vec{C}_{\text{LA}}\,, \nonumber \\ 
   \vec{I}^{\text{con}}_{\text{LB}} &= - \frac{\mu_{\text{LB}}}{2\pi} K_{0}^{-1}  \vec{t}_0 + R_{K_0}^{-1} P_{1} \vec{C}_{\text{LB}} \,, \nonumber \\ \vec{I}^{\text{con}}_{\text{RA}} &= \frac{\mu_{\text{RA}}}{2\pi} K_{0}^{-1} \vec{t}_0 +  R_{K_0}^{-1} P_1 \vec{C}_{\text{RA}}\,, \nonumber \\ \vec{I}^{\text{con}}_{\text{RB}} &= - \frac{\mu_{\text{RB}}}{2\pi} K_{0}^{-1}  \vec{t}_0 + R_{K_0}^{-1} P_{1/3} \vec{C}_{\text{RB}} \,.
\label{eq:I-con}
\end{align}
We have used the same sign convention as above for $\vec{I}_{\ell}^{\text{jun}}$. 
Here, $\mu_{\ell}$ with $\ell = \text{LA}, \text{LB}, \text{RA}, \text{RB}$ denote the chemical potentials of the respective contacts,
$ R_{K_0} = \text{diag} (1, \sqrt{3})$,  and $P_1$ ($P_{1/3}$) is a projection operator on the $\nu = 1$ (respectively, 1/3) mode, i.e., $P_{1} = (1 + \sigma_z)/2$ and $P_{1/3} = (1 - \sigma_z)/2$ in the basis of 1 and 1/3 modes that we use. 
In each of the formulas \eqref{eq:I-con}, the first term corresponds to the mode propagating out 
of the contact, which is thus equilibrated with the chemical potential of the contact. The second term corresponds to the mode going in the contact, so that the corresponding coefficients $\vec{C}_{\ell}$ are not specified by Eq.~\eqref{eq:I-con}.

We assume now that the arm segments are clean, so that the transport there is ballistic. (We will relax this condition in Sec.~\ref{sec:mesoscopicfluctuation}.) 
In this case, each individual component of currents is conserved, i.e., 
\begin{align}
  \vec{I}_{\ell}^{\text{jun}} =  \vec{I}^{\text{con}}_\ell\,,
  \label{eq:I-ballistic}
\end{align}
with $\ell = \text{LA}, \text{LB}, \text{RA}, \text{RB}$. Applying $R_{K_0} P_1$ ($R_{K_0} P_{1/3}$) to the equations for the arms in  region A (respectively, B), we obtain four equations 
\begin{align} \label{eq:equations}
    \mu_{\text{LA}} P_1 R_{K_0} K^{-1}_{0} \vec{t}= 
 P_1 R_{K_0} K^{-1}_{0} \vec{\mu}_{\text{LA}}^{\text{jun}}\,, \nonumber \\ 
    \mu_{\text{LB}} P_{1/3} R_{K_0} K^{-1}_{0} \vec{t}= 
 P_{1/3} R_{K_0} K^{-1}_{0} \vec{\mu}_{\text{LB}}^{\text{jun}}\,, \nonumber \\ 
 \mu_{\text{RA}} P_{1/3} R_{K_0} K^{-1}_{0} \vec{t}= 
 P_{1/3} R_{K_0} K^{-1}_{0} \vec{\mu}_{\text{RA}}^{\text{jun}}\,, \nonumber \\ 
 \mu_{\text{RB}} P_{1} R_{K_0} K^{-1}_{0} \vec{t}= 
 P_{1} R_{K_0} K^{-1}_{0} \vec{\mu}_{\text{RB}}^{\text{jun}}\,. 
\end{align}
Substituting here $\vec{\mu}_{\ell}^{\text{jun}}$ from Eq.~\eqref{eq:chemicalpotentialconst}, we obtain a system of four linear equations for four variables $\mu^{\rm red}_{a}$ (with $a=1,2$) and $\mu^{\rm loc}_{i}$ (with $i=R,L$). 

For each contact $\ell$ with $\ell = \text{LA}, \text{LB}, \text{RA}, \text{RB}$,
we denote by $J_{\ell}$ the corresponding total charge current. The sign convention here is such that $J_{\ell} > 0$ if the current flows to the contact and $J_{\ell} < 0$ for current flowing out of the contact. We have 
\begin{eqnarray}
J_{\ell} &=& -\vec{t}_0^T \cdot \vec{I}_{\ell}^{\text{jun}} = -\vec{t}_0^T \cdot \vec{I}^{\text{con}}_\ell \,, \qquad \ell = \text{LA, LB} \,, \nonumber \\
J_{\ell} &=& \vec{t}_0^T \cdot \vec{I}_{\ell}^{\text{jun}} = \vec{t}_0^T \cdot \vec{I}^{\text{con}}_\ell \,, \qquad \ell = \text{RA, RB} \,,
\label{eq:continuity}
\end{eqnarray}
where we used, in the last expressions, the condition \eqref{eq:I-ballistic} of the ballistic transport in the arms. The minus sign in the first line of Eq.~\eqref{eq:continuity} is related to our sign conventions.

For each of our two null vectors $\vec{M}$, we solve  Eqs.~\eqref{eq:equations} and \eqref{eq:chemicalpotentialconst}, and then use Eq.~\eqref{eq:continuity} to determine the conductance matrix $\mathcal{G}_{\ell \ell'}$ that connects currents $J_{\ell}$ with the applied bias voltages $\mu_{\ell'}$,
\begin{align} 
\label{eq:conductancematrixform}
   \begin{pmatrix}
        J_{\text{LA}} \\  J_{\text{RA}} \\  J_{\text{RB}} \\ J_{\text{LB}}
    \end{pmatrix}  
    = \mathcal{G}
    \begin{pmatrix}
        \mu_{\text{LA}} \\  \mu_{\text{RA}} \\  \mu_{\text{RB}} \\ \mu_{\text{LB}}
    \end{pmatrix}\,.
\end{align}
We find that $\mathcal{G}$ [in units of $e^2 / h = 1/ (2 \pi)$] generically takes the block matrix form as 
\begin{align} 
\label{eq:conductancematrix}
  \mathcal{G} = \begin{pmatrix}
        \mathcal{G}_0 &   \mathcal{G}_{\text{off}} [g_1]\\  \mathcal{G}_{\text{off}} [g_2] & \mathcal{G}_0
    \end{pmatrix}\,,
\end{align}
with 
\begin{align} \label{eq:conductancematrix2}
  \mathcal{G}_0 = \begin{pmatrix}
        - (1- g_0)  &   1/3 - g_0 \\ (1- g_0) & -(1/3 - g_0)
    \end{pmatrix}, \quad  \mathcal{G}_{\text{off}} [g] = \begin{pmatrix}
        g  &   -g \\ -g & g
    \end{pmatrix}, 
\end{align}
Note that the conductance matrix is determined by three real numbers $g_0$, $g_1$, and $g_2$, which parameterize the diagonal and off-diagonal blocks, respectively. The diagonal blocks $\mathcal{G}_0$ describe the current induced in the same edge to which the voltage is applied, while the off-diagonal blocks  $\mathcal{G}_{\text{off}} [g]$ 
characterize the current induced in the other edge. Since there is no tunneling between the edges, the current in each of the edges is conserved separately, i.e., $J_{\text{LA}} = - J_{\text{RA}}$ and $J_{\text{LB}} = - J_{\text{RB}}$, which is evident in the form of the conductance matrix, Eqs.~\eqref{eq:conductancematrix} and \eqref{eq:conductancematrix2}. We recall that the minus sign here comes from our sign convention: $J_\ell > 0$ corresponds to currents flowing into the contact for all $\ell$.

For the case of the localization driven by the null vector $\vec{M}_{\text{sup}}$, we obtain 
\begin{align} 
\label{eq:cleanneutralsuper}
    g_0 = g_1 = g_2 = \frac{1}{4}\,. 
\end{align}
On the other hand, for the case of localization via the $\vec{M}_{\text{back}}$ channel, 
we obtain instead
\begin{align} 
\label{eq:g0g1gapphase}
     g_0 = -g_1 = -g_2 =  \frac{1}{4}\,. 
\end{align}
Thus, the current induced in the other edge has the same magnitude but the opposite sign in comparison to the $\vec{M}_{\text{sup}}$ localization. 

\begin{figure}[t!]
\includegraphics[width =\columnwidth]{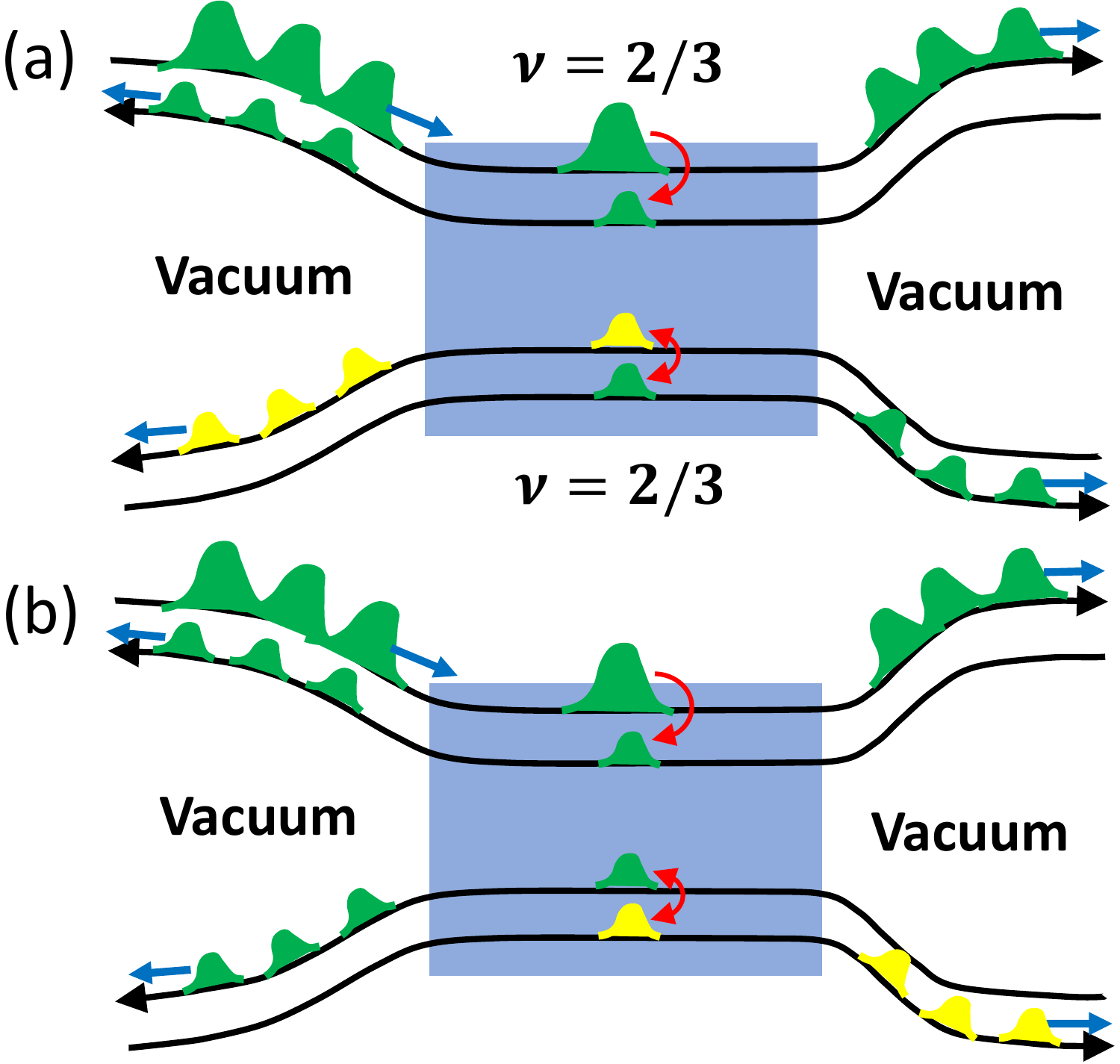}
\caption{
{\bf Drag induced by neutral-mode localization} at the junction (depicted as shaded blue region) for the localization in the (a) $\vec{M}_{\text{sup}}$ channel and (b) $\vec{M}_{\text{back}}$ channel.
 Bias is applied to  the upper left arm (via the LA contact, see Fig.~\ref{Fig:Setup}). Incoming electrons (depicted in green) from this arm are partially reflected at the junction. At the same time, a neutral excitation in the bottom edge (edge B) is created in accordance with the respective null vector, $\vec{M}_{\text{sup}}$ for (a) and $\vec{M}_{\text{back}}$ for (b). 
 This induces an electron-like excitation (green) and a hole-like excitation (yellow) in two modes of the edge B, which propagate in opposite directions, and arrive at different drains.
 Thus, a potential applied to the electrode LA induces a current in the edge B, amounting to a drag. The sign of the drag is positive for $\vec{M}_{\text{sup}}$ localization and negative for $\vec{M}_{\text{back}}$ localization. The effect can also be interpreted as a non-local Andreev reflection. 
} 
\label{Fig:Process}
\end{figure}

Note that $g_1=g_2$ for the localization in any of the channels, 
Eqs.~\eqref{eq:cleanneutralsuper}
and \eqref{eq:g0g1gapphase}. We 
introduced distinct parameters $g_1$ and $g_2$
in Eq.~\eqref{eq:conductancematrix} because below,
in Sec.~\ref{sec:mesoscopicfluctuation}, where coherent random tunneling in the arms is included, we will encounter situations with $g_1 \ne g_2$.

From the four-terminal conductance matrix, one can also obtain the value of the two-terminal conductance studied in Sec.~\ref{sec:twoterminal}. This is done by setting  $\mu_{\text{LA}} =  \mu_{\text{LB}} =  \mu_{\text{L}}$ and 
$\mu_{\text{RA}} =  \mu_{\text{RB}} =  \mu_{\text{R}}$, and calculating the total current from left to right, $J = J_{\text{RA}} + J_{\text{RB}}$. Using Eqs.~\eqref{eq:conductancematrix} and \eqref{eq:conductancematrix2} for the matrix $\mathcal{G}$, we get
\begin{equation}
G = \frac43 - 2g_0 + g_1 + g_2 \,.
\end{equation}
Substituting here the values of $g_0$, $g_1$ and $g_2$ for each of the localization channels,
Eqs.~\eqref{eq:cleanneutralsuper} and
\eqref{eq:g0g1gapphase}, we reproduce the results \eqref{eq:G-teo-term-sup-back} for the two-terminal conductance. In Table~\ref{tab:transport}, we summarize the results for the two-terminal conductance and for elements of the four-terminal conductance matrix for both neutral-mode localization channels.

\begin{table} 
\begin{tabular}{|c|| c | c | c |}
\hline
$\vec{M}$ & $G$ & \multicolumn{2}{|c|}{$\mathcal{G}_{\ell \ell'}$} \\ \hline
\multirow{2}{*}{$\vec{M}_{\text{sup}}$}  & \multirow{2}{*}{$4/3 $} & $\mathcal{G}_{\text{RA},\text{LA}} =  3/4$  & $\mathcal{G}_{\text{RB}, \text{LB}} =  1/12$ \\ & &  $\mathcal{G}_{\text{LB}, \text{LA}} = -1/4$ & $\mathcal{G}_{\text{LA}, \text{LB}} =  -1/4$
\\ \hline
\multirow{2}{*}{$\vec{M}_{\text{back}}$}     & \multirow{2}{*}{$1/3$} & $\mathcal{G}_{\text{RA}, \text{LA}} =  3/4$  & $\mathcal{G}_{\text{RB},\text{LB}} =  1/12$
\\ & & $\mathcal{G}_{\text{LB}, \text{LA}}  = 1/4$ & $\mathcal{G}_{\text{LA}, \text{LB}} =  1/4$ \\ \hline
\end{tabular}
\caption{{\bf Two-terminal conductance $G$ and four-terminal conductance matrix $\mathcal{G}_{\ell \ell'}$} (all in units of $e^2 /h$) in the case of clean (ballistic) arms, for  either superconducting neutral-mode localization (null vector $\vec{M}_{\text{sup}}$) or neutral back-scattering localization (null vector $\vec{M}_{\text{back}}$) in the junction. 
An element $\mathcal{G}_{\ell \ell'}$ of the four-terminal conductance matrix is defined as $J_\ell / \mu_{\ell'}$, where $J_\ell$ is the current from the junction into the  contact $\ell$ in response to the electrochemical potential $\mu_{\ell'}$ of contact $\ell'$, see Eq.~\eqref{eq:conductancematrixform}. Other elements of the conductance matrix follow from the current conservation in each of the edges, implying $\mathcal{G}_{\text{LA}, \ell'} = - \mathcal{G}_{\text{RA}, \ell'}$ and $\mathcal{G}_{\text{LB}, \ell'} = - \mathcal{G}_{\text{RB}, \ell'}$, and from the symmetry of the device with respect to rotation by 180$^{\circ}$ (i.e., the 
transformation LA $\leftrightarrow$ RB and
RA $\leftrightarrow$ LB). }
\label{tab:transport}
\end{table}

The mechanism of generation of the current in edge B when the voltage is applied to one of the edge-A contacts is illustrated in Fig.~\ref{Fig:Process}. 
The tunneling process corresponding to the null vector $\vec M$ that governs localization (i.e., either $\vec{M}_{\text{sup}}$ or $\vec{M}_{\text{back}}$) involves a back-scattering in edge A and simultaneously creation of a neutral excitation in edge B. The electron and hole components of this neutral excitations (one of them in the mode 1 and another in the mode 1/3) propagate in opposite directions and are collected in different contacts (LB and RB), thus implying a current induced in the edge B. 

This phenomenon bears analogy to the Coulomb drag effect that happens in double-layer electronic systems, see Ref.~\cite{Narozhny2016May} for a review. In most of the cases, the Coulomb drag is weak and explicitly depends on the interaction strength. In our case, the drag blocks $\mathcal{G}_{\text{off}}$ of the conductance matrix are of order unity (i.e., of the same order as the blocks $\mathcal{G}_0$ corresponding to the conductance within the edge). Furthermore, the drag conductance has a quantized value $ \pm 1/4$.
This is related to the fact that we consider not a perturbative regime but rather a strong localization of neutral modes. In this connection, there is a certain similarity with FQH bilayers with total filling factor $\nu=1$, where exciton condensation may take place, leading to strong and quantized drag \cite{Kellogg2002observation,Nandi2012exciton,Eisenstein2014exciton}. There are, however, many obvious differences between the system that we consider and the $\nu=1$ bilayers. In particular, we study here the edge physics, while the exciton condensation in bilayers is a bulk effect. 

In the last few years, a number of remarkable experiments on transport in FQH edge junctions with counterpropagating modes were interpreted in terms of Andreev scattering 
\cite{Hashisaka2021,Cohen2023,Glidic2023quasiparticle}. Conventionally, the term ``Andreev scattering'' is used to describe scattering in SN or SNS structures (with ``S'' and ``N'' denoting superconductor and normal metal, respectively). In this case, the total charge of excitations is not conserved (since a Cooper pair can go to the condensate). As a result, an electron can be back-scattered as a hole, which is the Andreev scattering. In FQH experiments mentioned above there is no superconductor involved, so that the total charge of excitations is conserved. The term ``Andreev scattering'' has been used in a broader sense in this context, describing situations when, say, a positive charge impinging on a device induces a negative charge flowing from the device to one of the contacts (charge conservation is maintained by charges propagating to other contacts).  The transport in junctions with neutral-mode localization that we consider can be also interpreted in this fashion. Indeed, consider, e.g., the case of localization in the $\vec{M}_{\text{sup}}$ channel, with a bias applied to the LA electrode and other contacts grounded. Then electrons impinging on the junction from the LA arm produce hole-like excitations propagating from the junction into the contact LB,
see Fig.~\ref{Fig:Process}(a).  For the case of localization in the $\vec{M}_{\text{back}}$ channel, the hole-like excitations move to contact RB, see Fig.~\ref{Fig:Process}(b).

We finally consider the four-terminal conductance in the case of fully equilibrated arms. 
As was discussed in Sec.~\ref{sec:twoterminal},
the effect of incoherent equilibration in each of the arms can be effectively taken into account by replacing $R_{K_0}$ with $U = O R_{K_0}$ 
in Eq.~\eqref{eq:equations}, where $O \in $ SO(1,1) is  \begin{align}
    O =  \begin{pmatrix} \cosh \vartheta & \sinh \vartheta  \\ \sinh \vartheta & \cosh \vartheta \end{pmatrix} \,. 
\end{align}
We find $\vartheta = \frac{1}{2} \log ( 2+ \sqrt{3})$ in the fully equilibrated limit of $\gamma_{ab} \gg \Gamma_a$, see Eq.~\eqref{eq:equalibrationmatrix}. By solving Eqs.~\eqref{eq:equations} and \eqref{eq:chemicalpotentialconst} with this replacement, and then using Eq.~\eqref{eq:continuity}, we obtain
\begin{align} \label{eq:fourtermalfullequilibration}
    g_0 = \frac{1}{3}\,, \quad g_1 = g_2 = 0\,. 
\end{align}
Thus, neutral-mode localization does not affect the four-terminal conductance matrix  in the regime of equilibrated transport in the arms. 
In particular, the two-terminal conductance is $G=2/3$ [as was already found above, see Eq.~\eqref{eq:supercondequilibration}]
and the drag is absent. 

It is worth noting that the drag should remain non-zero in the partially equilibrated regime (i.e., in the crossover between the ballistic and equilibrated regimes). In this regime, $|g_1|$ and $|g_2|$ are expected to be between $0$ (Eq.~\eqref{eq:fourtermalfullequilibration}) and $\frac{1}{4}$ (Eqs.~\eqref{eq:cleanneutralsuper} and \eqref{eq:g0g1gapphase}), interpolating between the two limiting transport regimes, fully equilibrated and ballistic.

\section{Mesoscopic fluctuations} \label{sec:mesoscopicfluctuation}

In Sec.~\ref{sec:fourterminalcond}, we determined the four-terminal conductance matrix for each of the neutral-mode localization channels by assuming that the transport in the arms is ballistics (i.e., the arms are clean). If the arms are disordered and the transport there is incoherent, neutral-mode localization does not affect the transport, see Sec.~\ref{sec:twoterminal}. In this section, we consider a regime that is intermediate between these two limits, namely there is random tunneling in the arms but transport is coherent. This is the case when the 
length of the arms is larger than the zero-temperature mean-free path and, at the same time, the temperature is low enough, so that the inelastic equilibration length is larger than the arm length. In this regime, 
the interaction in the random edge is renormalized towards the Kane-Fisher-Polchinski fixed point  at which eigenmodes diagonalizing the interaction matrix are the charge and neutral modes \cite{Kane_Randomness_1994}. Furthermore,
transport properties in this regime depend on realization of disorder, implying mesoscopic fluctuations of conductances \cite{Protopopov_transport_2_3_2017, Rosenow_signatures_2010}.

In the mesoscopic regime, a specific realization of random tunneling in a 2/3 edge segment (representing any of the arms of our setup) affects transport through total phase factor $e^{i\theta}$, i.e., via a single parameter  $\theta \in[0,2\pi]$. If the modes of such a segment are directly connected to contacts,  the transport is characterized by  a $2\times 2$ matrix
 \cite{Protopopov_transport_2_3_2017}
 \begin{align} 
 \label{eq:condmat}
   \mathcal{G}_{\text{arm}} =  \begin{pmatrix}
        1- g_{\text{arm}} & g_{\text{arm}} \\ g_{\text{arm}} & \frac{1}{3} -g_{\text{arm}}
    \end{pmatrix},
\end{align}
with an only back-scattering parameter $g_{\text{arm}}$ which should satisfy $0\le g_{\text{arm}} \le 1/2 $. 
The matrix \eqref{eq:condmat} connects the outgoing currents (i.e., those flowing to the leads) in the 1 and 1/3 modes with the potentials applied to incoming 1 and 1/3 modes. 
Combining two such arms, one obtains a two-terminal conductance $G = 4/3 - g_{\text{arm}}^{\text{t}} - g_{\text{arm}}^{\text{b}}$, with superscripts ``t'' and ``b'' for top and bottom arms. When the disorder is modified, the conductance thus experiences mesoscopic fluctuations in the range $1/3 \le G \le 4/3$.  This mesoscopic regime was observed in a recent experiment \cite{Hashisaka2023}. 
Analytical calculation of the conductances for a generic $\theta \in[0,2\pi]$ is a highly challenging task since the bosonized action is no more quadratic in this situation. The analysis becomes simpler in the cases $\theta=0$ (which is the same as no disorder, $g_{\text{arm}}=0$) and $\theta=\pi$, for which one finds the largest possible value $g_{\text{arm}}=1/2$. 
For $\theta=\pi$, the bosonic action remains quadratic but the boundary condition is modified: the sign of the neutral mode is flipped.
If one includes in consideration sufficiently long contact regions (``leads'') where 1 and 1/3 modes are non-interacting, the parameter $\theta$ gets renormalized, with $\theta=0$ and $\theta=\pi$ being stable and unstable fixed points, respectively \cite{Protopopov_transport_2_3_2017}.

We extend now the calculation of four-terminal conductance of Sec.~\ref{sec:fourterminalcond} by allowing for random tunneling in the arms. We consider first the case when all $\theta_\ell$ are either 0 or $\pi$, so that an exact calculation is possible, and then discuss what happens for other values.

Restricting ourselves to either $\theta_\ell = 0$ or $\theta_\ell = \pi$ for each of the arms, $\ell = \text{LA, RA, LB, RB}$, we get 16 possible combinations. We have evaluated the four-terminal conductance matrix for all of them.  The random tunneling with  $\theta_\ell = \pi$ on an arm $\ell$ leads to a modification of the analysis of boundary matching of currents (Sec.~\ref{sec:fourterminalcond}) by a flip of sign of the neutral bosonic mode in the corresponding arm.  
We find that, for both localization channels, the conductance matrix retains its form, Eqs.~\eqref{eq:conductancematrix} and \eqref{eq:conductancematrix2}, for all 16 cases.
The parameters $g_0$, $g_1$, $g_2$ get, however, affected by random tunneling in the arms. Specifically, we obtain
\begin{align} 
\label{eq:conductancecoh}
\left \{ \begin{array}{llll}
 g_0 = g_1 = g_2 = \frac{1}{4} & \,\, \text{for}\,\, \theta_{\text{LA}} = \theta_{\text{LB}}\,\, \text{and} \,\, \theta_{\text{RA}} = \theta_{\text{RB}} \,,
 \\ [1mm]  g_0 = \minus g_1 = \minus g_2 = \frac{1}{4} &\,\,   \text{for} \ 
\theta_{\text{LA}} \neq \theta_{\text{LB}}\,\, \text{and} \,\, \theta_{\text{RA}} \neq \theta_{\text{RB}}\,, 
 \\  [1mm]
 g_0 = \frac{2}{5}, \,\, g_1 = \minus g_2 = \frac{1}{5}  &\,\,  \text{for} \ 
\theta_{\text{LA}} = \theta_{\text{LB}}\,\, \text{and} \,\, \theta_{\text{RA}} \neq \theta_{\text{RB}}\,, 
 \\  [1mm]
g_0 = \frac{2}{5}, \,\,  g_2 = \minus g_1 =  \frac{1}{5}  &\,\,  \text{for} \ 
\theta_{\text{LA}} \neq \theta_{\text{LB}}\,\, \text{and} \,\, \theta_{\text{RA}} = \theta_{\text{RB}}\, 
 \end{array} \right .
\end{align}
for the case of localization in the junction via the $\vec{M}_{\text{sup}}$ channel and
\begin{align} 
\label{eq:conductancecoh-2}
\left \{ \begin{array}{llll}
 g_0 = \minus g_1 = \minus g_2 = \frac{1}{4} & \,\, \text{for}\,\, \theta_{\text{LA}} = \theta_{\text{LB}}\,\, \text{and} \,\, \theta_{\text{RA}} = \theta_{\text{RB}} \,,
 \\ [1mm]  g_0 =  g_1 =  g_2 = \frac{1}{4} &\,\,   \text{for} \ 
\theta_{\text{LA}} \neq \theta_{\text{LB}}\,\, \text{and} \,\, \theta_{\text{RA}} \neq \theta_{\text{RB}}\,, 
 \\  [1mm]
 g_0 = \frac{2}{5}, \,\, g_2 =  \minus  g_1 = \frac{1}{5}  &\,\,  \text{for} \ 
\theta_{\text{LA}} = \theta_{\text{LB}}\,\, \text{and} \,\, \theta_{\text{RA}} \neq \theta_{\text{RB}}\,, 
 \\  [1mm]
g_0 = \frac{2}{5}, \,\, g_1 = \minus g_2 =  \frac{1}{5}  &\,\,  \text{for} \ 
\theta_{\text{LA}} \neq \theta_{\text{LB}}\,\, \text{and} \,\, \theta_{\text{RA}} = \theta_{\text{RB}}\, 
 \end{array} \right .
\end{align}
for the case of $\vec{M}_{\text{back}}$ localization.

The first and second lines of Eqs.~\eqref{eq:conductancecoh} 
and \eqref{eq:conductancecoh-2}
can be understood by inspecting the condition of localization in channel $\vec M$,
Eq.~\eqref{eq:loc-condition-M}, which has the form
 $\vec{M}^T \vec{I} (x) = 0$ on each side (left and right) of the junction, or, writing explicitly A and B components, 
 \begin{equation}
 \vec{M}_{\text{A}}^T \vec{I}_{\text{A}} (x) +  \vec{M}_{\text{B}}^T \vec{I}_{\text{B}} (x) = 0\,.
 \label{eq:loc-condition-AB}
 \end{equation}
When expressed in the charge-neutral basis, the condition  \eqref{eq:loc-condition-AB} has the form
\begin{equation}
    \vec{M}^T_{\text{A}} \vec{I}_{n \text{A}} (x) + \vec{M}^T_{\text{B}} \vec{I}_{n \text{B}} (x) = 0 \,,
\label{eq:loc-condition-AB-n}
    \end{equation}
since the localization considered here takes place in the neutral sector, as expressed by Eq.~\eqref{eq:neutral-loc}. 
 As shown in Ref.~\cite{Protopopov_transport_2_3_2017} and discussed above, $\theta = \pi$  
 in an edge segment leads to a flip of the sign
 (i.e., the $\pi$ phase shift) of the neutral bosonic mode, $\phi_n \rightarrow - \phi_n$ and hence $\vec{I}_{n} (x) \rightarrow - \vec{I}_{n} (x)$. For the $\vec{M}_{\text{sup}}$ localization and clean arms (i.e., $\theta_\ell = 0$ for all $\ell$), we had the values $g_0 = g_1 = g_2 = 1/4$, see 
 Eq.~\eqref{eq:cleanneutralsuper}. 
 This result holds more generally for
 $\theta_{\text{LA}} = \theta_{\text{LB}}$ and $\theta_{\text{RA}} = \theta_{\text{RB}}$,
 see the first line of Eq.~\eqref{eq:conductancecoh},  since the 
 condition \eqref{eq:loc-condition-AB-n}
remains the same when the signs of both currents
$\vec{I}_{n \text{A}}$  and $\vec{I}_{n \text{B}}$ are flipped. The same arguments explain the first line of Eq.~\eqref{eq:conductancecoh-2} for the case of $\vec{M}_{\text{back}}$ localization.
For $\theta_{\text{LA}} \neq \theta_{\text{LB}}$ and $\theta_{\text{RA}} \neq  \theta_{\text{RB}}$, either $\vec{I}_{n\text{A}}$ or $\vec{I}_{n\text{B}}$ changes sign on each side (L and R) of the junction. Upon such a transformation, the condition \eqref{eq:loc-condition-AB-n} becomes
\begin{equation}
    \vec{M}^T_{\text{A}} \vec{I}_{n \text{A}} (x) - \vec{M}^T_{\text{B}} \vec{I}_{n \text{B}} (x) = 0 \,,
\label{eq:loc-condition-AB-n-sign-flip}
\end{equation}
i.e., the vector $\vec M^T = (\vec{M}^T_{\text{A}}, \vec{M}^T_{\text{B}})$ is mapped onto 
$\widetilde{\vec M}^T = (\vec{M}^T_{\text{A}}, - \vec{M}^T_{\text{B}})$. This maps $\vec{M}_{\text{sup}}$ onto $\vec{M}_{\text{back}}$
and vice versa, thus explaining the second lines of Eqs.~\eqref{eq:conductancecoh} and \eqref{eq:conductancecoh-2}: The conductances for 
$\vec{M}_{\text{sup}}$ localization with
arms characterized by 
$\theta_{\text{LA}} \neq \theta_{\text{LB}}$ and $\theta_{\text{RA}} \neq  \theta_{\text{RB}}$
are the same as those for $\vec{M}_{\text{back}}$ localization with clean arms, and vice versa. 

The last two lines of
Eqs.~\eqref{eq:conductancecoh} and \eqref{eq:conductancecoh-2} represent a more tricky scenario:  The condition on one side (e.g., left) of the junction retains its form \eqref{eq:loc-condition-AB-n} while on the other side (correspondingly, right) of the junction the sign is flipped, Eq.~\eqref{eq:loc-condition-AB-n-sign-flip}. In other words, on one side, the condition is similar to a system with ballistic arms and  $\vec{M}_{\text{sup}}$ localization, while on the other side it is similar to a system with ballistic arms and  $\vec{M}_{\text{back}}$ localization.
This explains the emergence of values $(g_0, g_1, g_2)$ different from those in the cases of localization in any of the two localization channels with clean arms.  It is further instructive to inspect 
implications of the 180$^\circ$ rotation transformation (which exchanges $\text{LA} \leftrightarrow \text{RB}$ and $\text{RA} \leftrightarrow \text{LB}$).
Since this transformation interchanges the left and right sides, the conditions in the third and fourth lines of Eq.~\eqref{eq:conductancecoh} 
are exchanged, and analogously for 
Eq.~\eqref{eq:conductancecoh-2}. At the same time, this transformation interchanges the parameters $g_1 \leftrightarrow g_2$ of the conductance matrix, Eqs.~\eqref{eq:conductancematrix} and \eqref{eq:conductancematrix2}. We see that the results for the parameters $g_i$ in the last two lines of Eqs.~\eqref{eq:conductancecoh}, \eqref{eq:conductancecoh-2} are in agreement with this symmetry: $g_1$ in the third line is equal to $g_2$ in the fourth line and vice versa.

In general, if the disorder in the arms is continuously changed, 
the values of $\theta_\ell$ continuously change as well. As a result, the conductance matrix will exhibit mesosocopic fluctuations, interpolating between its values calculated above for $\theta_\ell$ equal to 0 or $\pi$. This will in particular imply mesoscopic fluctuations of the drag, including both its magnitude and the sign. 
Specifically, by analyzing general properties of the four-terminal conductance matrix, we find that the drag conductances $g_1$ and $g_2$ vary in the range from $-\frac{1}{4}$ to $\frac{1}{4}$, see Appendix~\ref{app:dragconductancerange} for details. Thus, the first two lines in
Eqs.~\eqref{eq:conductancecoh} and \eqref{eq:conductancecoh-2} correspond to extreme values of the drag conductances.

As discussed above, the effect of random tunneling in the arms can be reduced if a setup with engineered 2/3 edges (as realized in Refs.~\cite{Cohen2019, Hashisaka2021, Hashisaka2023})
is used, see Fig.~\ref{Fig:Setup}(c).

\section{Summary}
\label{sec:summary}

We have explored manifestations of Anderson localization of neutral modes in transport properties of a junction of two 2/3 FQH edges. Localization in one of the two competing channels (``neutral-mode superconductivity'' and ``neutral mode backscattering'', governed by null vectors $\vec{M}_{\text{sup}}$ and $\vec{M}_{\text{back}}$, respectively) reduces the original four-mode theory of the junction to an effective two-mode theory. This effective theory has a bosonic character: The minimal integer-charge excitations are bosons with charge two, and elementary fractional excitations have charge 2/3 (i.e., twice larger than the charge 1/3 in the absence of localization). These values of the charge can be probed in tunneling experiments.

We have determined the two-terminal and four-terminal conductances of a junction with neutral-mode localization. For the case of ballistic transport in the arms connecting the junction to contacts, the two-terminal conductance is $G=4/3$  for $\vec{M}_{\text{sup}}$ localization and $G=1/3$ for $\vec{M}_{\text{back}}$ localization. The four-terminal conductance provides more information on transport through the junction. In particular, it reveals the drag phenomenon: A bias applied to edge A induces a transport in edge B, even though there is no tunneling of charge between the edges. The drag conductances are quantized and equal to $\pm 1/4 (e^2/h)$; the sign of the drag is opposite in two localization  channels. This phenomenon implies, in particular, that electrons approaching the junction from one of the contacts induce hole-like excitations propagate from the junction to another contact, which can be viewed as a special type of non-local Andreev scattering. 

We have also studied the effect of coherent random tunneling in arms of the device (which are segments of 2/3 edges) on the four-terminal conductance matrix. Such random tunneling leads to strong mesoscopic fluctuations of the conductances. Interestingly, the random tunneling may effectively transform one of the neutral localization channels into the other one.

We emphasize that Anderson localization is a coherent phenomenon, so that its experimental observation in FQH devices is certainly a challenging task. At the same time, recent years have witnessed remarkable advances in FQH device engineering and experimental studies of quantum-coherent transport in FQH regime. Thus, experimental detection and investigation of localization in FQH edges and edge junctions---including the neutral-mode localization studied theoretically in this paper---appears to be feasible. We hope that the present work will stimulate experimental research in this direction, which may be expected to provide further important insights into remarkably rich physics of FQH quantum matter. 

\appendix

\section{Range of drag conductance}
\label{app:dragconductancerange}

In this appendix, we determine the range of possible values of the drag conductances $g_1$ and $g_2$ and of the two-terminal conductance $G$. By employing energy conservation argument, as in Refs.~\cite{Wen1994, Sen2008}, we show that 
$g_1$ and $g_2$ are bounded to the range $[-\frac{1}{4}, \frac{1}{4}]$; at the same time, $G$ is limited to the interval $[\frac{1}{3}, \frac{4}{3}]$. 

We consider a generic form of the four-terminal conductance matrix $\mathcal{G}$, parameterized by four parameters $g_1$, $g_2$, $g_\text{A}$, and $g_{\text{B}}$, 
\begin{align}
\label{eq:conductancematrixapp}
  \mathcal{G} = \begin{pmatrix}
        \mathcal{G}_0 [g_{\text{A}}] &   \mathcal{G}_{\text{off}} [g_1]\\  \mathcal{G}_{\text{off}} [g_2] & \mathcal{G}_0 [g_{\text{B}}]
    \end{pmatrix}\,,
\end{align}
with 
\begin{align} 
\label{eq:conductancematrix2app}
  \mathcal{G}_0 [g] = \begin{pmatrix}
        - (1- g)  &   1/3 - g \\ (1- g) & -(1/3 - g)
    \end{pmatrix}, 
\end{align}
and $\mathcal{G}_{\text{off}}[g]$ given by Eq.~\eqref{eq:conductancematrix2}.
This conductance matrix generalizes Eq.~\eqref{eq:conductancematrix} that includes three parameters $g_1$, $g_2$, and $ g_{\text{A}} = g_{\text{B}} = g_0$. To motivate
Eq.~\eqref{eq:conductancematrixapp}, we 
note that $\mathcal{G}$ is a $4 \times 4$ matrix and thus has 16 matrix elements. Current conservation in each edge (i.e., $J_{\text{LA}} = - J_{\text{RA}}$ and  $J_{\text{LB}} = - J_{\text{RB}}$) implies that the first and second rows of $\mathcal{G}_0 [g]$ and of $\mathcal{G}_{\text{off}}[g]$ are identical up to an opposite sign, yielding eight constraints. Furthermore, for $\mu_{\text{LA}} = \mu_{\text{LB}} = \mu_{\text{RA}} = \mu_{\text{RB}} = \mu$, the currents in the arms should be equal to their equilibrium values, $J_{\text{RA}} = J_{\text{LB}}  = - J_{\text{LA}} =  - J_{\text{RB}} = \frac{2}{3} \big (\frac{\mu}{2\pi} \big)$. It thus follows from $J_{\ell} = \frac{1}{2\pi} \sum_{\ell'} \mathcal{G}_{\ell \ell'} \mu_{\ell'} =  \frac{\mu}{2\pi} (\sum_{\ell'} \mathcal{G}_{\ell \ell'})$ that the sum 
$\sum_{\ell'} \mathcal{G}_{\ell \ell'}$ of the components in row $\ell$ is fixed to $-2/3$ for $\ell = \text{LA}, \text{RB}$ (and correspondingly to 2/3 for two other rows). This yields further two constraints. 
Finally, when each edge is separately in equilibrium,  i.e., $\mu_{\text{LA}} = \mu_{\text{RA}}$ and $\mu_{\text{LB}} = \mu_{\text{RB}}$, neutral-mode localization is not operative, since the corresponding tunneling operator \eqref{eq:tunnelinglocalization} does not couple to the total charge density of each edge, and thus the drag should be absent. This leads to two additional constraints that the sum of two columns of each of the off-diagonal (drag) blocks  of $\mathcal{G}$ vanishes.
These latter constraints imply that the off-diagonal blocks have the form \eqref{eq:conductancematrix2}, as has been verified in all the cases for which we calculated $\mathcal{G}$ explicitly in Sec.~\ref{sec:mesoscopicfluctuation}. 
Since we have in total $8 + 2 +2 = 12$ independent constraints, the matrix $\mathcal{G}$ is expressed in terms of four parameters, which explains
 Eq.~\eqref{eq:conductancematrixapp}.

\begin{figure}[t!]
\includegraphics[width =\columnwidth]{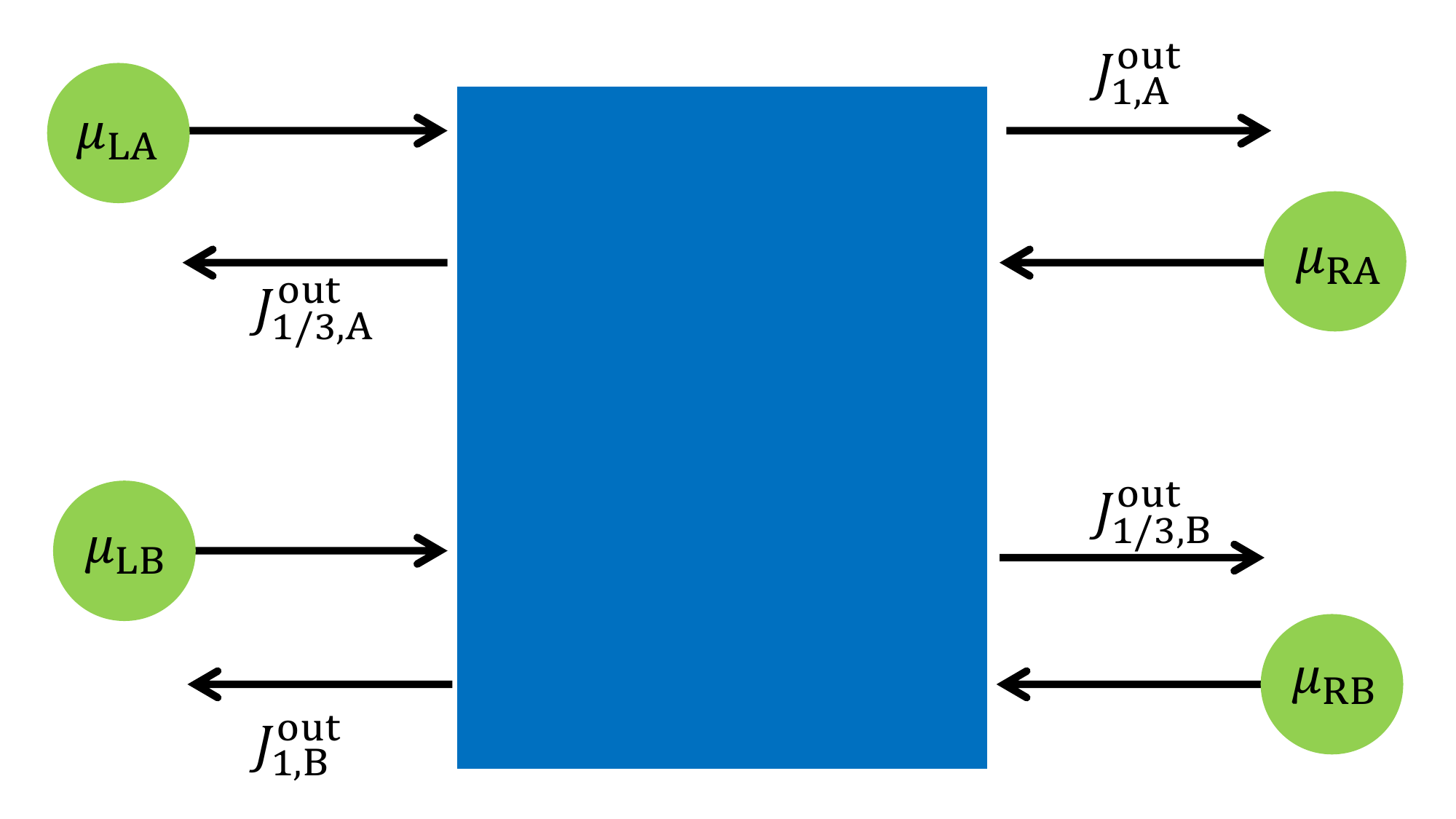}
\caption{
{\bf Generic four-terminal setup} with four incoming  incoming and four outgoing modes.
The blue region is a ``black box'' that includes both the junction and the arms in Fig.~\ref{Fig:Setup}(b). 
The currents in outgoing modes, $J_{1,\text{A}}^{\text{out}}$,  $J_{1,\text{B}}^{\text{out}}$,  $J_{1/3,\text{A}}^{\text{out}}$, and $J_{1/3,\text{B}}^{\text{out}}$ are indicated, with subscripts referring to the edge (A or B) and the type of the mode (1 or 1/3). 
} 
\label{Fig:Blackbox}
\end{figure}

We consider a generic four-terminal setup depicted in Fig.~\ref{Fig:Blackbox}. Here, the whole system (i.e., the junction together with the arms in Fig.~\ref{Fig:Setup}(b)) is represented as one 
``black box'', with four incoming and four outgoing modes. The idea \cite{Wen1994, Sen2008} is to calculate the incoming and outgoing power carried by the currents. The total incoming power (energy per unit time) reads 
\begin{align}
    P_{\text{in}} = \frac{1}{4 \pi} (\mu_{\text{LA}}^2 + \mu_{\text{RB}}^2 + \frac{1}{3}\mu_{\text{RA}}^2 + \frac{1}{3}\mu_{\text{LB}}^2 ) \,.  
\end{align}
On the other hand, the total power associated with outgoing currents is given by 
\begin{align}
       P_{\text{out}} = \pi [(J_{1,\text{A}}^{\text{out}})^2 + (J_{1,\text{B}}^{\text{out}})^2  + 3 (J_{1/3,\text{A}}^{\text{out}})^2 + 3 (J_{1/3,\text{B}}^{\text{out}})^2]\,. 
\end{align}
By using the conductance matrix \eqref{eq:conductancematrixapp} and the relations between currents 
\begin{align}
    J_{\text{LA}} = J_{1/3, \text{A}}^{\text{out}} - \frac{\mu_{\text{LA}}}{2\pi}\,,& \quad 
    J_{\text{RA}} = J_{1, \text{A}}^{\text{out}} - \frac{1}{3} \frac{\mu_{\text{RA}}}{2\pi}\,, \nonumber \\ 
    J_{\text{RB}} = J_{1/3, \text{B}}^{\text{out}} - \frac{\mu_{\text{RB}}}{2\pi}\,,& \quad 
    J_{\text{LB}} = J_{1, \text{B}}^{\text{out}} -  \frac{1}{3}  \frac{\mu_{\text{LB}}}{2\pi}\,, 
\end{align}
we obtain the power difference, $\Delta P = P_{\text{in}}- P_{\text{out}}$, expressed as a quadratic form 
\begin{align}
    \Delta P  = \frac{1}{4\pi} \vec{\mu}^T \widetilde{\mathcal{G}} \vec{\mu}\,,
\end{align}
with respect to the vector or chemical potentials, $\vec{\mu}^T = (\mu_{\text{LA}},\mu_{\text{RA}},\mu_{\text{RB}},\mu_{\text{LB}} )$. 
Here, the matrix $\widetilde{\mathcal{G}}$ is found to be
\begin{align}\label{eq:conductancematrixapp2}
    \widetilde{\mathcal{G}} 
  = \begin{pmatrix}
       \widetilde{\mathcal{G}}_0 [g_{\text{A}}, g_2] &    \widetilde{\mathcal{G}}_{\text{off}}\\   \widetilde{\mathcal{G}}_{\text{off}} &  \widetilde{\mathcal{G}}_0 [g_{\text{B}}, g_1]
    \end{pmatrix}\,,
\end{align}
with 
\begin{align}\label{eq:conductancematrixapp22}
    \widetilde{\mathcal{G}}_0 [g, g']
 =  [ -4 g'^2 + 2 (1- 2 g)g ] \begin{pmatrix}
      1 &  -1 \\   -1 &  1
    \end{pmatrix}\,, \nonumber \\ 
      \widetilde{\mathcal{G}}_{\text{off}} = 
       [g_1 + g_2 - 4 (g_2 g_{\text{B}} + g_1 g_{\text{A}})]
      \begin{pmatrix}
       1 &    -1\\   -1 &   1
    \end{pmatrix}\,. 
\end{align}
The energy difference  $\Delta P$ is dissipated (transformed into heat) in scattering processes in the system. Thus, $\Delta P \geq 0$. This inequality should hold for any choice of $\vec{\mu}$, and hence the matrix $\widetilde{\mathcal{G}}$ should be positive semi-definite, which imply constraints on possible values of the parameters $g_1$, $g_2$, $g_\text{A}$, and $g_{\text{B}}$. In particular,
the positive semi-definiteness of $\widetilde{\mathcal{G}}$ results in the following constraints on drag conductances
\begin{align}
\label{eq:drag-bounds}
    -\frac{1}{4} \leq g_1, g_2 \leq \frac{1}{4}\,. 
\end{align}
Further, the two-terminal conductance $G = \frac{4}{3} - g_{\text{A}} -g_{\text{B}} + g_1 + g_2$ is bounded to the range $[\frac{1}{3}, \frac{4}{3}]$, which is the same as in the absence of any coupling between two 2/3 edges. 

The upper and lower bounds \eqref{eq:drag-bounds} on $g_1$ and $g_2$,
exactly correspond to the first two lines of Eqs.~\eqref{eq:conductancecoh} and \eqref{eq:conductancecoh-2}, which means that they are the optimal bounds. Thus, the drag conductances vary within the range \eqref{eq:drag-bounds} in the regime of mesoscopic fluctuations, Sec.~\ref{sec:mesoscopicfluctuation}.

\begin{acknowledgments}
J.P., Y.G., and A.D.M. acknowledge support by the DFG Grant MI 658/10-2. 
Y.G. and J.I.V. acknowledge support by grant no 2022391 from the United States - Israel Binational Science Foundation
(BSF), Jerusalem, Israel. 
M.G. has been supported by the Israel Science Foundation (ISF) and the Directorate for Defense
Research and Development (DDR\&D) Grant No. 3427/21, the ISF grant No. 1113/23, and the BSF Grant No. 2020072. Y. G. is supported by  the Minerva foundation. Y. G. is an incumbent of InfoSys chair.
\end{acknowledgments}

%

\end{document}